\documentclass[10pt,twocolumn,twoside,a4paper]{extarticle}

\usepackage[a4paper,top=3.4cm,bottom=3.4cm,left=1.9cm,right=1.9cm,columnsep=0.65cm]{geometry}

\usepackage[T1]{fontenc}
\usepackage{textcomp}
\usepackage{mathptmx}                 
\usepackage[scaled=0.92]{helvet}      
\usepackage{microtype}

\usepackage{graphicx}
\graphicspath{{figures/}}
\usepackage{amsmath,amssymb}
\usepackage{bm}
\usepackage{booktabs}
\usepackage{multirow}
\usepackage{array}
\usepackage{tabularx}
\usepackage{xcolor}

\usepackage{fancyhdr}
\usepackage{titlesec}
\usepackage[font=footnotesize,labelfont={bf,sf},textfont={sf},labelsep=space,
            justification=justified,singlelinecheck=false,skip=5pt]{caption}
\usepackage{lastpage}
\usepackage{marvosym}                 
\usepackage[switch]{lineno}
\usepackage[hidelinks]{hyperref}

\definecolor{lsablue}{RGB}{46,73,124}
\definecolor{lsarule}{RGB}{120,150,190}
\definecolor{lsagray}{RGB}{90,90,90}

\fancypagestyle{firstpage}{\fancyhf{}%
  \fancyhead[R]{\footnotesize Page \thepage\ of \pageref{LastPage}}}

\titleformat{\section}{\sffamily\bfseries\fontsize{11}{13}\selectfont}{}{0pt}{}
\titlespacing{\section}{0pt}{10pt plus 2pt minus 1pt}{3pt}
\titleformat{\subsection}{\sffamily\bfseries\fontsize{10}{12}\selectfont}{}{0pt}{}
\titlespacing{\subsection}{0pt}{8pt plus 2pt minus 1pt}{2pt}

\renewcommand{\arraystretch}{1.15}

\usepackage[super,numbers,sort&compress]{natbib}
\providecolor{docnotelinkcolor}{rgb}{0,0,0}  

\newenvironment{lsaabstract}
  {\vspace{11pt}\par\noindent
   {\sffamily\bfseries\fontsize{11}{13}\selectfont Abstract}\par\vspace{3pt}
   \begingroup\sffamily\fontsize{10}{12}\selectfont\noindent\ignorespaces}
  {\par\endgroup\vspace{3pt}}

\newcommand{\bmsize}{\fontsize{8}{10}\selectfont}
\newcommand{\bmhead}[1]{\par\addvspace{9pt}\noindent
  {\sffamily\bfseries\fontsize{8}{10}\selectfont #1}\par\nobreak\vspace{2pt}}

\begin{document}

\twocolumn[
\begin{@twocolumnfalse}
\vspace*{6pt}

{\fontsize{19}{23}\selectfont
Bias-resolved comb-formation regimes in a simple two-section quantum-dot mode-locked laser for isolator-free, terabit-scale O-band interconnects\par}

\vspace{10pt}

{\fontsize{10.5}{13}\selectfont
Ying Shi, William He, Xiangpeng Ou, Defan Sun, Weiping Li, Xin Yao and Yating Wan\textsuperscript{\Letter}\par}

\begin{lsaabstract}
Practical O-band quantum-dot (QD) mode-locked comb sources must combine wide flat-top bandwidth, low radio-frequency (RF) timing noise, error-correctable bare-line modulation, and feedback tolerance, yet these functions are usually optimized separately and their relation to comb-formation dynamics remains unclear. Here we map the gain-current/saturable-absorber-bias plane of a simple two-section InAs/GaAs QD mode-locked laser to resolve distinct application-specific regimes. A low-injection, high-reverse-bias regime produces 0.80 ps pulses, consistent with amplitude-modulated operation, whereas a high-injection, intermediate-bias regime yields a 16.16 nm (2.75 THz) flat-top comb with 110 lines and a 14.98 ps extended waveform, consistent with a stronger frequency-modulated contribution. To our knowledge, this is the broadest reported 3 dB bandwidth among QD mode-locked comb sources. At a separate low-noise bias, the beatnote exhibits a 0.58 kHz Lorentzian linewidth and a 41.6~fs integrated timing jitter over the 4 to 80~MHz range, to our knowledge the lowest reported for a high-channel-count O-band QD passive comb. In isolator-free 25~Gb/s NRZ transmission, all 94 carriers across a 13.73 nm 3 dB band remain below the 7\% HD-FEC threshold; using all 110 carriers at the maximum-bandwidth bias gives a projected aggregate rate of 2.75~Tb/s. At a separate feedback-test bias, broadband mode locking is preserved and, beyond approximately $-$28~dB feedback, the RF beatnote enters a feedback-stabilized regime whose linewidth is reduced 42-fold. Thus, short-pulse, broadband, low-noise, transmission-ready, and feedback-stabilized operation occupy distinct selectable regions rather than a single universal optimum. This regime-resolved map links QD comb-formation physics to isolator-free, terabit-scale O-band interconnects.
\end{lsaabstract}

\vspace{6pt}
\end{@twocolumnfalse}
]

\thispagestyle{firstpage}

\makeatletter
\renewcommand\@makefntext[1]{\noindent#1}
\makeatother
\renewcommand{\thefootnote}{}
\footnotetext{\fontsize{7.6}{9.5}\selectfont
Correspondence: Yating Wan (\href{mailto:yating.wan@kaust.edu.sa}{yating.wan@kaust.edu.sa})\\
Integrated Photonics Lab, King Abdullah University of Science and Technology (KAUST), Thuwal 23955-6900, Kingdom of Saudi Arabia}
\renewcommand{\thefootnote}{\arabic{footnote}}

\section{Introduction}

Practical O-band wavelength-parallel links require a compact, electrically driven source that can supply many near-uniform carriers while maintaining low radio-frequency (RF) timing noise, error-correctable per-line modulation, and tolerance to residual reflections so that a discrete optical isolator is unnecessary. Optical frequency combs (OFCs) are attractive for this role because one chip-scale source can replace an array of discrete lasers~\cite{diddams2020nature,hermans2022APL}, while supporting a broad range of applications in optical communications~\cite{marin-palomo2017nature,bernal2024nc,hu2021Nanophotonics,wang2025eLight}, optical computing~\cite{feldmann2021nature,nie2024optica,zhou2026AP}, microwave photonics~\cite{liu2020np}, spectroscopy~\cite{villares2014NC,van2020arXiv}, and light detection and ranging~\cite{trocha2022SR,yao2024Elight,yang2024ELight}.

Among chip-scale comb technologies, semiconductor mode-locked lasers (MLLs) are distinctive in being compact, directly electrically driven, and free of an external optical pump or high-Q passive resonator~\cite{senica2026nature,davenport2018pr,liang2022jstqe,he2024SiPhotonics,liu2023jlt,sun2023oe,zhou2023elight}. For an interconnect comb, however, optical bandwidth alone is insufficient: the bias that broadens the spectrum need not shorten the pulse, minimize timing diffusion, or maximize feedback tolerance. A single optimized spectrum or operating point therefore cannot establish the full capability of a semiconductor-comb platform.

Quantum-dot (QD) gain media are particularly promising for O-band MLLs because their three-dimensional carrier confinement maps naturally onto these requirements. Inhomogeneously broadened gain spectrum supports wide optical bandwidth and high comb-line count~\cite{rafailov2007np,norman2019JQE}; a small linewidth enhancement factor and reduced spontaneous-emission noise suppress RF timing noise~\cite{duan2018APL}; fast carrier and saturable absorption dynamics support passive mode locking~\cite{borri2000jstqe}; weak feedback sensitivity enables isolator-free operation~\cite{shi2026lsa,cui2024APL,ou2026optica}, alongside low threshold current~\cite{wan2017optica,tan2025Light}, high-temperature stability~\cite{kageyama2011extremely,wan2018PR}, and long device lifetime~\cite{shang2021optica}.

Importantly, the same gain-absorber system can support distinct comb states rather than one universal optimum. Gain dynamics, spatial hole burning, group-velocity dispersion, Kerr nonlinearity, and four-wave mixing can shift the balance between amplitude-modulated (AM-like) pulse formation and frequency-modulated (FM-like) broadband operation~\cite{hillbrand2020prl,opacak2019prl,dong2023lsa,wang2025ap,roy2024optica,prokoshin2026APR}; delayed optical feedback can further reorganize the phase and timing dynamics. The relevant physical question is therefore which comb-formation and feedback regimes provide the different functions required by a wavelength-parallel source.

Recent O-band QD combs have advanced individual figures of merit: a chirped 20~GHz device demonstrated a 6.1 nm 3~dB bandwidth with 58 lines, a 1.8 kHz RF linewidth, 82.7 fs timing jitter, and 4.1~Tb/s transmission capacity~\cite{liu2019optica}; harmonic colliding-pulse devices achieved $>$11 nm flat-top 100~GHz combs with high-speed PAM transmission~\cite{huang2022pr,huang2024pr}; commercial QD combs delivered $>$2 THz usable bandwidth and high output power~\cite{buyalo2024sr,rautert2025ofc}; and a co-doped 100~GHz device achieved 14.31~nm bandwidth with 26 lines~\cite{pan2025lpr} (Table~\ref{tab:benchmark}). These results establish the strength of the QD platform, but the leading metrics have often been obtained using additional design degrees of freedom, such as chirped stacks, higher-harmonic colliding-pulse cavities, or co-doping, and bandwidth has been reported under non-uniform criteria. This makes it difficult to separate platform-level capability from architecture-specific enhancement. Moreover, short-pulse, broadband, low-noise, transmission, and feedback performance are commonly characterized in different devices or at isolated bias points. For 100~GHz-class combs, the fundamental beatnote also lies beyond the direct range of standard electrical spectrum analysis, limiting quantitative access to RF linewidth, phase noise, and feedback dynamics. Record comparisons alone therefore do not reveal how these functions are related within one QD MLL.

Here we address this question using a deliberately simple two-section InAs/GaAs QD MLL as a controlled platform for bias-resolved measurements. The narrow saturable absorber (SA) maintains a small absorber saturation energy and efficient intensity-dependent loss, whereas the tapered gain section increases the gain saturation energy and supports power extraction while preserving the gain-absorber contrast required for passive mode locking. The architecture itself is not claimed as new. Rather, avoiding chirped stacks, harmonic cavity geometry, and co-doping reduces additional structural degrees of freedom, while the 25~GHz fundamental beatnote remains directly accessible to electrical spectrum analysis. We map the full gain-current/SA-bias plane and correlate optical bandwidth, pulse width, RF coherence, per-line transmission, and controlled-feedback response in the same physical device.

The maps resolve distinct application-specific regimes. A low-injection, high-reverse-bias regime produces 0.80~ps pulses, consistent with AM-like operation. By contrast, a high-injection, intermediate-bias regime produces a flat-top comb with a 3~dB optical bandwidth of 16.16~nm (2.75~THz) and 110 lines, together with a temporally extended waveform of 14.98~ps effective width, consistent with a stronger FM-like contribution. To our knowledge, this is the broadest reported 3~dB bandwidth among QD mode-locked comb sources. At a low-noise bias, the fundamental beatnote exhibits a 0.58 kHz Lorentzian RF linewidth and a 41.6~fs integrated timing jitter over the ITU-T-specified 4 to 80~MHz range, to our knowledge the lowest reported value for a high-channel-count O-band QD passive comb.

At a separate transmission bias, all 94 measured carriers across a 13.73 nm 3 dB band sustain isolator-free 25~Gb/s NRZ transmission below the 7\% HD-FEC threshold; applying the measured per-carrier rate to the 110 carriers available at the maximum-bandwidth bias gives a projected aggregate line rate of 2.75~Tb/s. At another feedback-test bias within the broadband-comb region, the optical envelope and 25~GHz temporal structure remain preserved under controlled delayed reflection. Beyond approximately $-28$~dB feedback, the RF beatnote enters a feedback-stabilized regime and narrows from 190~kHz to 4.50~kHz, approximately a 42-fold linewidth reduction. Thus, the same simple QD MLL accesses short-pulse, broadband, low-noise, transmission-ready, and feedback-stabilized functions across distinct selectable regions rather than at a single universal optimum. This regime-resolved operating map connects QD comb-formation and feedback dynamics to isolator-free, terabit-scale O-band interconnects and provides a quantitative baseline for higher-repetition-rate QD comb sources.

\begin{figure*}[!t]
\centering
\includegraphics[width=\textwidth]{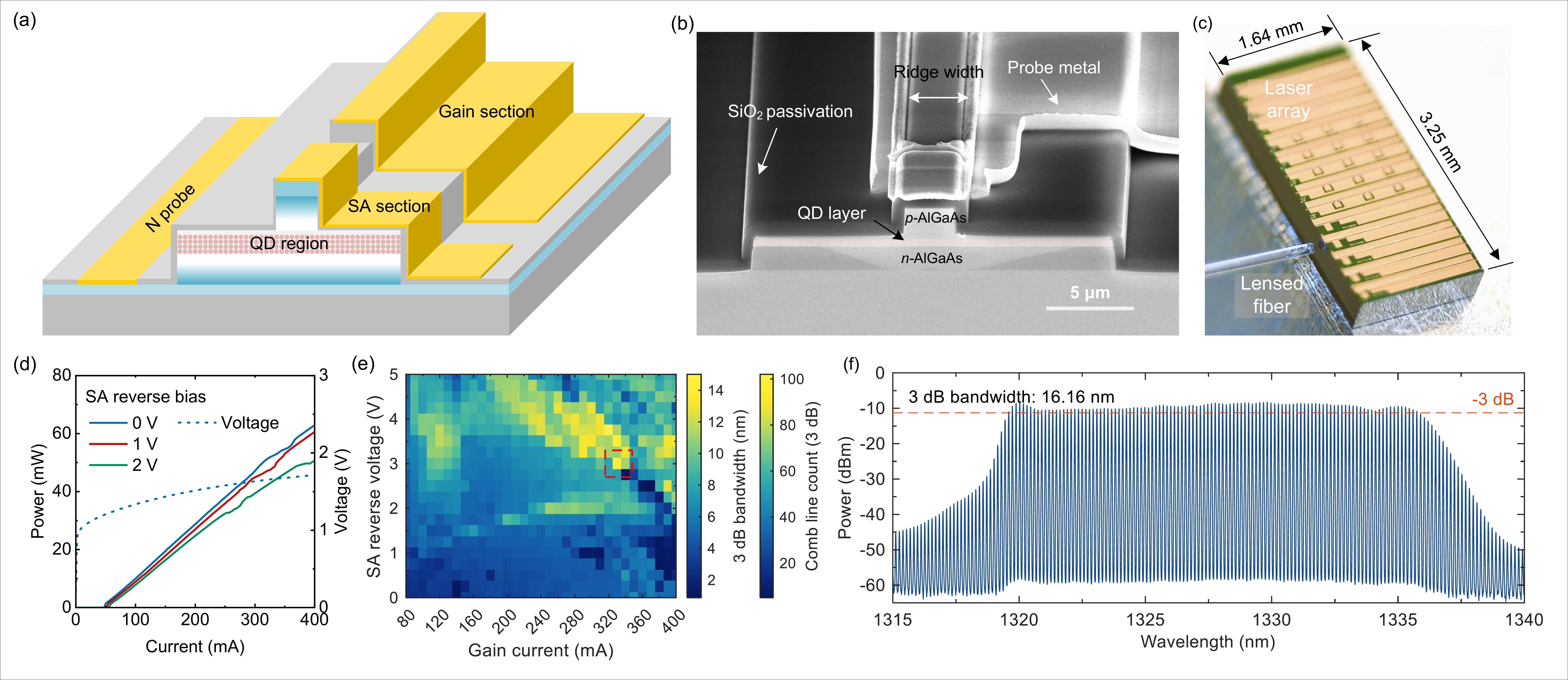}
\caption{\textbf{Device structure and broadband comb generation of the 25~GHz QD mode-locked laser.} \textbf{a} Schematic of the shallow-etched two-section Fabry-Perot cavity with a tapered gain section and a reverse-biased saturable absorber (SA). \textbf{b} Cross-sectional SEM image of the ridge waveguide, showing the QD active region, SiO$_2$ passivation, and probe metal. \textbf{c} Optical micrograph of the laser array under lensed-fiber coupling, with thirteen devices on a 1.64~mm~$\times$~3.25~mm bar. \textbf{d} Light-current-voltage characteristics at representative SA reverse biases, showing a threshold current of 48~mA at $V_\mathrm{SA}=0$~V. \textbf{e} Bias-dependent 3~dB optical bandwidth map, with the secondary color scale indicating the corresponding comb-line count. The star marks the maximum-bandwidth point, $I_\mathrm{gain}=332$~mA and $V_\mathrm{SA}=-3.0$~V, identified by a refined bias scan. \textbf{f} Optical spectrum at this bias point, showing a flat-top comb centered near 1328~nm with a 3~dB bandwidth of 16.16~nm, corresponding to 2.75~THz and 110 comb lines.}
\label{fig:device}
\end{figure*}

\section{Results}

\subsection{Minimal two-section platform and broadband operating window}
\label{sec:device}

Figure~\ref{fig:device} establishes the deliberately minimal device and the broadband operating window used for the regime mapping. The device is a shallow-etched two-section Fabry-Perot cavity formed by a forward-biased tapered gain section and a reverse-biased saturable absorber (SA), shown schematically in Fig.~\ref{fig:device}a. The total cavity length is 1638~\textmu m, giving a fundamental repetition rate near 25~GHz for a group index of about 3.66. The SA is placed at one cleaved facet and occupies 14\% of the cavity length. Its ridge width is uniform at 3~\textmu m, while the gain-section ridge is linearly tapered from 3~\textmu m at the SA side to 6~\textmu m at the output facet. The fundamental-cavity design intentionally avoids harmonic colliding-pulse operation and keeps the RF beatnote directly measurable.

The tapered geometry provides the device-level saturation-energy balance that underpins bias-selectable operation. The reverse-biased SA provides intensity-dependent loss that promotes pulse shortening, whereas gain saturation in the forward-biased section tends to broaden the pulse. The narrow 3~\textmu m SA ridge keeps the modal area and absorber saturation energy small, preserving efficient pulse shortening. The widening gain-section ridge raises the gain-section saturation energy, suppressing gain-saturation-induced broadening and supporting higher output power without sacrificing the gain-absorber contrast required for stable passive mode locking. The shallow-etched ridge limits sidewall-roughness scattering loss while retaining adequate lateral confinement. Similar tapered QD MLL geometries have been explored as a route to combined short-pulse generation, high output power, and low-noise operation~\cite{thompson2009jstqe,bardella2018ol}. Here the geometry is used as a minimal baseline: gain current and SA bias tune the gain-absorber balance without introducing additional cavity sections or engineered harmonic operation.

The epitaxy was grown by molecular beam epitaxy on an \textit{n}-type GaAs substrate and comprises multiple stacked self-assembled InAs QD layers within a GaAs/AlGaAs waveguide and cladding. The ridge cross-section and layer stack are resolved in the SEM image of Fig.~\ref{fig:device}b. The full layer stack and detailed fabrication process flow are given in Supplementary Information. Figure~\ref{fig:device}c shows a fabricated bar carrying thirteen devices on a 1.64~mm $\times$ 3.25~mm footprint, coupled by a lensed fiber.

The DC isolation resistance between the gain and SA electrodes is about 3~k$\Omega$, allowing independent current injection and reverse-bias control. The light-current-voltage characteristics in Fig.~\ref{fig:device}d give a threshold current of 48~mA at $V_\mathrm{SA}=0$~V, a single-facet slope efficiency of $0.185$~W/A, and an output power exceeding 60~mW at $I_\mathrm{gain}=400$~mA. Reverse-biasing the SA adds intracavity loss, raising the threshold to 56~mA at $V_\mathrm{SA}=-2$~V and moderately lowering the slope efficiency. The measured series resistance of the device is below 2~$\Omega$.

To locate the broadest-comb regime, the laser was mapped over the gain-current and SA-reverse-bias plane ($I_\mathrm{gain}=80$ to $400$~mA, $V_\mathrm{SA}=0$ to $-5$~V) at $T=20\,^{\circ}\mathrm{C}$, recording the output spectrum at 0.02~nm resolution at each point. The resulting 3~dB optical-bandwidth map, with comb-line count on the secondary scale, is shown in Fig.~\ref{fig:device}e. Broadband operation is sustained over an extended, continuous region at intermediate reverse bias rather than at an isolated point, spanning a wide range of gain currents and indicating a robust operating window. The bandwidth peaks at $I_\mathrm{gain}=332$~mA and $V_\mathrm{SA}=-3.0$~V (star, located by a refined scan), where the spectrum in Fig.~\ref{fig:device}f is a flat-top envelope centered near 1328~nm with a 3~dB bandwidth of 16.16~nm, equivalent to 2.75~THz and 110 comb lines at the 25~GHz spacing. To our knowledge this is the broadest 3~dB bandwidth reported among QD mode-locked comb sources. The fact that the broadband state occupies a continuous branch rather than a fortuitous single point provides a reproducible regime for comparison with the time-domain and RF maps and for the subsequent transmission and feedback measurements.

\begin{figure*}[!t]
\centering
\includegraphics[width=\textwidth]{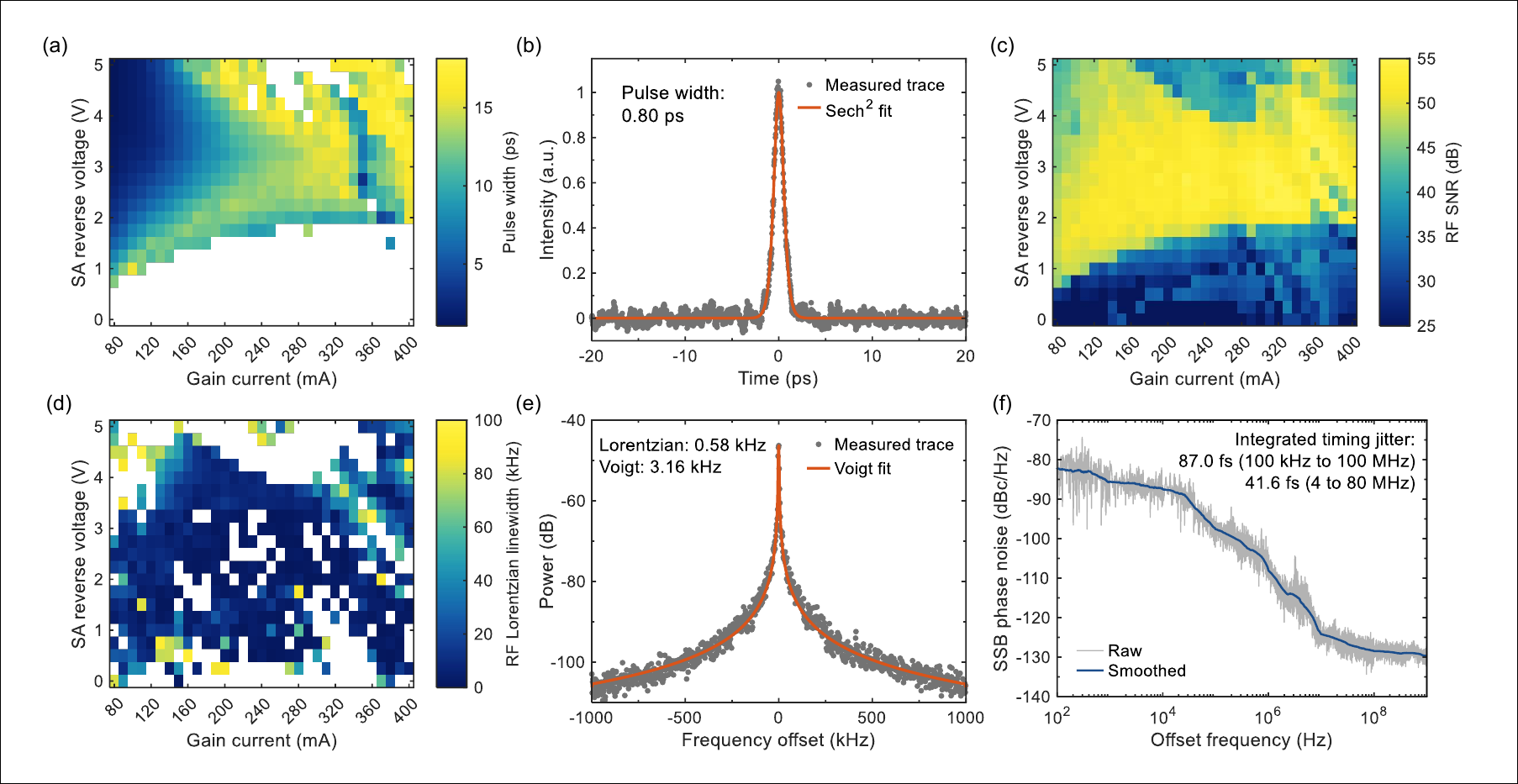}
\caption{\textbf{Pulse characteristics and low-noise RF performance of the 25~GHz QD mode-locked laser.} \textbf{a} Bias-dependent deconvolved pulse-width map extracted from intensity autocorrelation measurements. \textbf{b} Representative autocorrelation trace and $\mathrm{sech}^{2}$ fit at the shortest-pulse operating point, $I_\mathrm{gain}=80$~mA and $V_\mathrm{SA}=-5.0$~V, giving a pulse width of 0.80~ps. \textbf{c} RF signal-to-noise-ratio (SNR) map of the fundamental beatnote across the gain-current and SA-reverse-bias plane. \textbf{d} Lorentzian RF linewidth map from Voigt fits; white regions are excluded by the SNR or fit-quality criteria. \textbf{e} Narrow-span RF beatnote and Voigt fit at a representative low-noise point ($I_\mathrm{gain}=300$~mA, $V_\mathrm{SA}=-2.5$~V), giving a 0.58~kHz Lorentzian linewidth and a 3.16~kHz Voigt FWHM (RBW: 100 Hz). \textbf{f} Corresponding single-sideband phase noise of the fundamental beatnote. The integrated timing jitter is 87.0~fs from 100~kHz to 100~MHz and 41.6~fs from 4~MHz to 80~MHz.}
\label{fig:pulse_rf}
\end{figure*}

\begin{figure*}[!t]
\centering
\includegraphics[width=\textwidth]{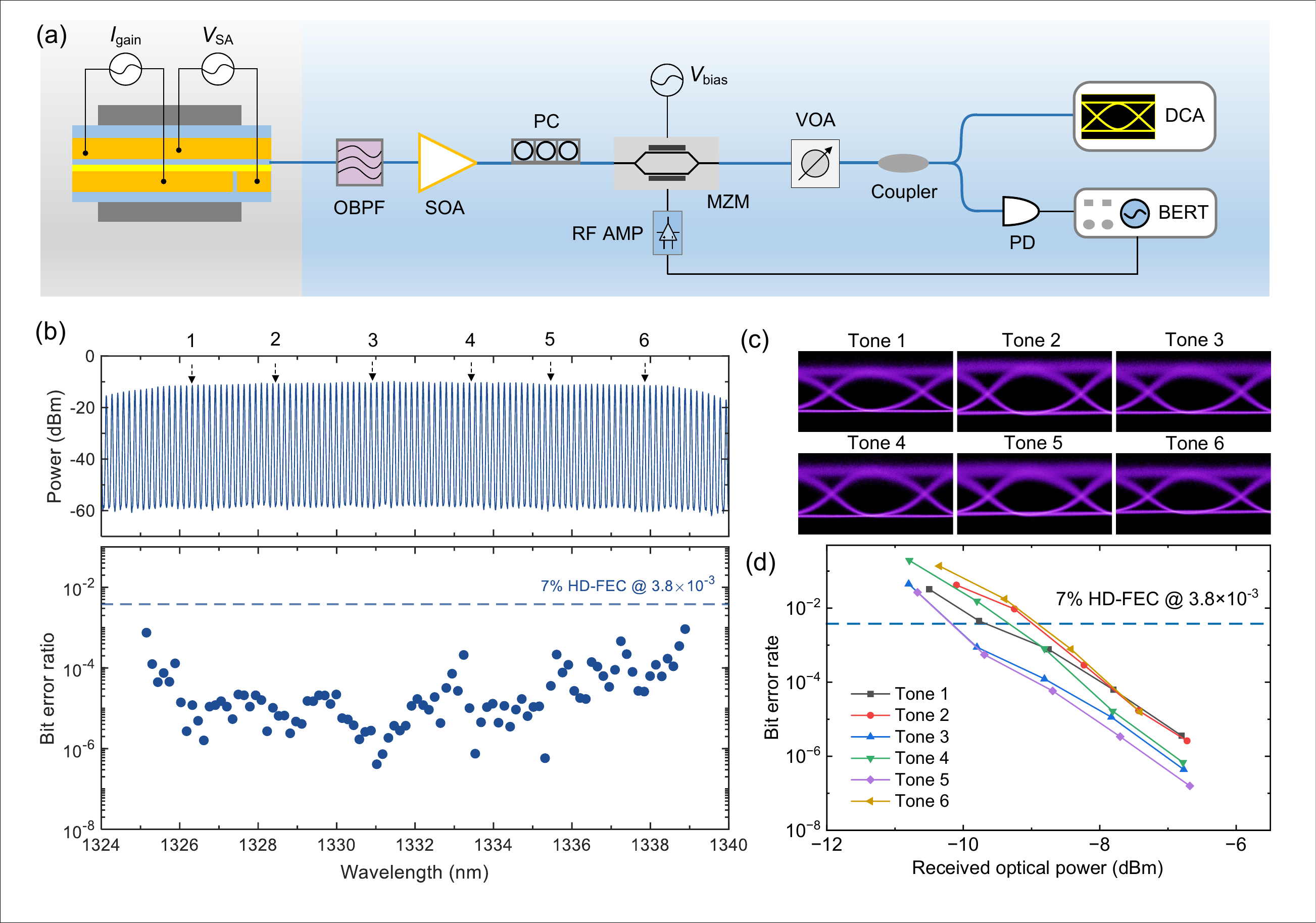}
\caption{\textbf{Isolator-free data transmission using the 25~GHz QD mode-locked laser.} \textbf{a} Experimental setup for isolator-free intensity-modulation/direct-detection (IM/DD) transmission: individual comb lines are selected by a tunable optical bandpass filter (OBPF), amplified by an O-band semiconductor optical amplifier (SOA), externally modulated by a Mach-Zehnder modulator (MZM), and detected for eye-diagram and bit-error-rate (BER) measurements. The MZM DC bias was occasionally adjusted slightly during the wavelength sweep to maintain near-quadrature operation. No optical isolator is placed between the laser facet and the transmission link. VOA, variable optical attenuator; PC, polarization controller; RF AMP, RF amplifier; BERT, bit-error-rate tester; DCA, digital communication analyzer; PD, photodiode; NRZ, non-return-to-zero; HD-FEC, hard-decision forward-error correction. \textbf{b} Comb-line spectrum used as wavelength carriers (upper) and per-channel BER across the 13.73~nm 3~dB bandwidth at 25~Gb/s NRZ (lower); all 94 measured channels stay below the 7\% HD-FEC threshold of $3.8\times 10^{-3}$ (dashed), each taken at the maximum received optical power of that channel ($-5.9$ to $-6.9$~dBm). Arrows mark the six representative channels (tone 1 to tone 6) characterized in \textbf{c} and \textbf{d}. \textbf{c} Back-to-back eye diagrams of representative channels under 25~Gb/s NRZ modulation. \textbf{d} BER versus received optical power for representative channels Tone 1 to Tone 6 distributed across the 3~dB bandwidth.}
\label{fig:transmission}
\end{figure*}

\subsection{Bias-resolved short-pulse, broadband, and low-noise regimes}
\label{sec:pulse_rf}

To resolve the time-domain branch of the operating plane, intensity autocorrelation traces were recorded at each operating point, and the pulse width was extracted by fitting the central peak with a $\mathrm{sech}^{2}$ profile and applying the corresponding deconvolution factor. The resulting pulse-width map, Fig.~\ref{fig:pulse_rf}a, shows the shortest pulses in the low-injection, high-reverse-bias corner, whereas the high-injection region that yields the broadest combs exhibits relatively broader autocorrelation traces. At the shortest-pulse bias of $I_\mathrm{gain}=80$~mA and $V_\mathrm{SA}=-5.0$~V, the autocorrelation in Fig.~\ref{fig:pulse_rf}b gives a minimum deconvolved pulse width of 0.80~ps.

At the maximum-bandwidth point ($I_\mathrm{gain}=332$~mA, $V_\mathrm{SA}=-3.0$~V), the central autocorrelation broadens to an effective width of 14.98~ps. The coexistence of a broad, spectrally flat envelope with a temporally extended waveform at high injection is consistent with a stronger frequency-modulated (FM-like) contribution to the comb dynamics, in which broad spectral coverage is accompanied by quasi-CW or temporally extended emission rather than short pulses. Comparable behavior has been discussed for semiconductor frequency combs, previously in quantum-cascade lasers and more recently in QD platforms, where gain dynamics, group-velocity dispersion, Kerr nonlinearity, and four-wave mixing jointly shape the transition between amplitude-modulated (AM-like) pulse formation and FM-like broadband operation~\cite{hillbrand2020prl,dong2023lsa,wang2025ap,opacak2019prl,roy2024optica,prokoshin2026APR}.

The short-pulse and broadband states therefore coexist within the same 25~GHz two-section QD MLL but occupy different regions of the bias plane. The low-injection, high-reverse-bias regime gives the cleanest sub-picosecond response for time-domain applications, whereas the high-injection, intermediate-reverse-bias region maximizes flat-top bandwidth and comb-line count for wavelength-parallel applications such as optical interconnects. The central design implication is that the conventional pulse-compression optimum is not the relevant optimum for a carrier-dense interconnect source.

The RF characteristics confirm stable fundamental mode locking over a broad operating region. RF coherence was surveyed over the same $(I_\mathrm{gain}, V_\mathrm{SA})$ plane used for the bandwidth and pulse maps, with the gain and SA sections driven by programmable low-noise sources to enable reproducible mapping. Residual technical noise from the drive electronics and environment can broaden the measured beatnote, so the linewidth map is a comparative indicator of RF coherence across the plane rather than the ultimate linewidth under fully optimized biasing. At each point, the narrow-span spectrum was fitted with a Voigt profile, whose Lorentzian component reflects the intrinsic timing-diffusion contribution while the full width additionally carries Gaussian technical and environmental broadening. The signal-to-noise ratio (SNR) map in Fig.~\ref{fig:pulse_rf}c shows an RF SNR above 40~dB across a large portion of the bias plane, including the broadband-comb region identified from the optical-bandwidth map. The Lorentzian linewidth map in Fig.~\ref{fig:pulse_rf}d shows sub-20~kHz linewidths over an extended middle-bias region. White regions correspond to points where stable mode locking was not obtained, the beatnote was unstable, or the fit did not meet the SNR and quality criteria.

At a representative low-noise point of $I_\mathrm{gain}=300$~mA and $V_\mathrm{SA}=-2.5$~V, the narrow-span beatnote is well described by the Voigt fit of Fig.~\ref{fig:pulse_rf}e, yielding a 0.58~kHz Lorentzian linewidth and a 3.16~kHz Voigt FWHM (full width at half maximum, equivalent to the $3$~dB width) at a 100~Hz resolution bandwidth. The sub-kilohertz Lorentzian component confirms highly stable passive mode locking. Integrating the single-sideband phase-noise spectrum of Fig.~\ref{fig:pulse_rf}f gives timing jitters of 87.0~fs over 100~kHz to 100~MHz and 41.6~fs from 4 to 80~MHz of the ITU-T-specified range. The integration method and the cumulative-jitter curve are detailed in Supplementary Information. To our knowledge, the 41.6~fs value is the lowest integrated timing jitter reported for a high-channel-count O-band QD passively mode-locked comb source, improving on the 82.7~fs of a 20~GHz QD MLL~\cite{liu2019optica}. This low timing noise is consistent with the reduced spontaneous-emission noise and small linewidth enhancement factor of the QD gain medium, with further suppression expected from improved packaging and shielding~\cite{carpintero2009PTL}.

Taken together, Figs.~\ref{fig:device}e and \ref{fig:pulse_rf}a to \ref{fig:pulse_rf}d resolve three experimentally distinguishable regions within one cavity: a short-pulse branch, a high-injection broadband branch, and a middle-bias low-noise branch. The regions partially overlap, so broad operation can remain coherent, but their extrema do not coincide. This regime separation, rather than a single optimized point, provides the operating logic for the transmission and feedback experiments below.

\begin{figure*}[!t]
\centering
\includegraphics[width=\textwidth]{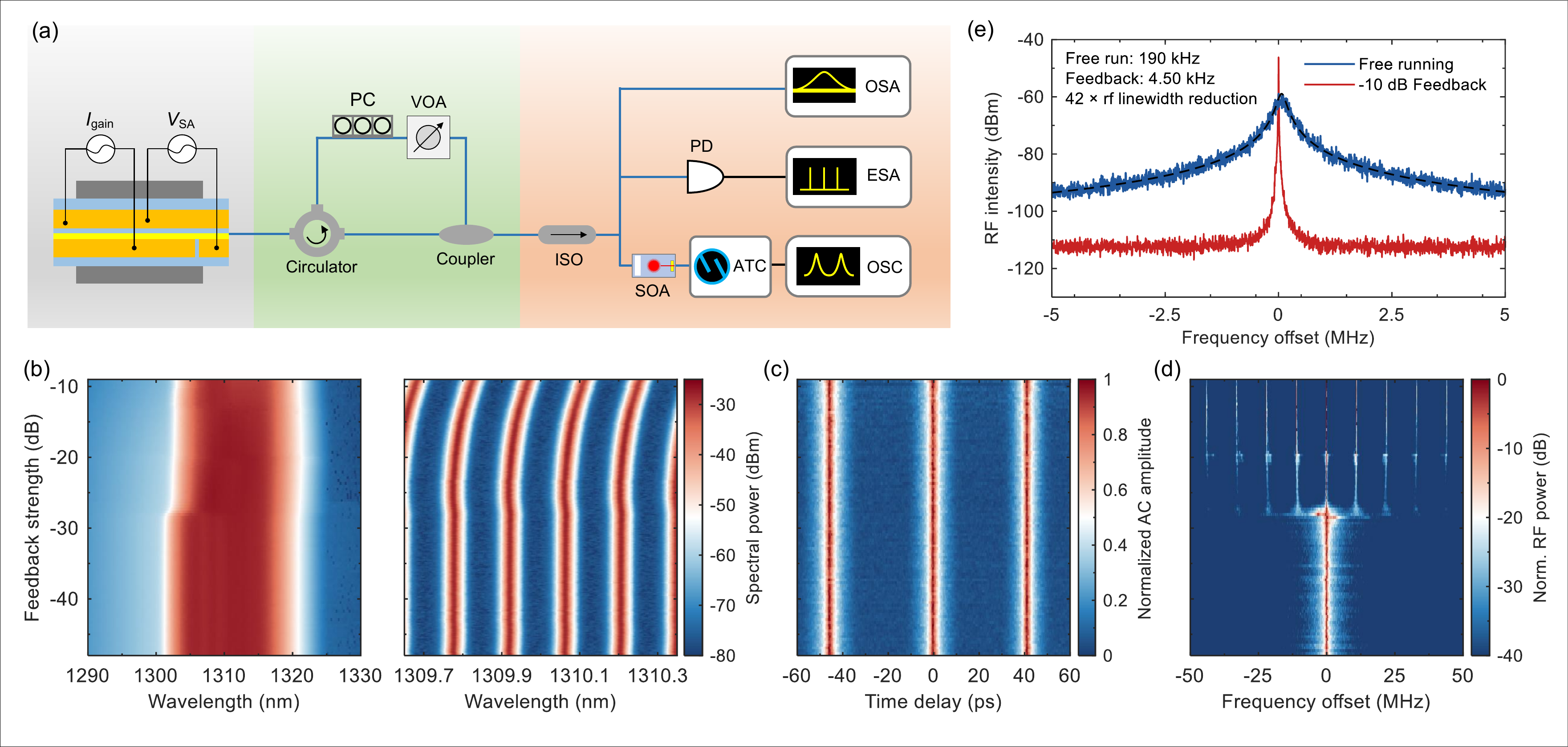}
\caption{\textbf{Controlled optical-feedback response of the 25~GHz QD mode-locked laser.} \textbf{a} Experimental setup: the output passes through a circulator into a fiber feedback path whose strength and polarization are set by a variable optical attenuator and polarization controller, while the optical spectrum, autocorrelation, time-domain waveform, and RF spectrum are monitored simultaneously. ISO, isolator; OSA, optical spectrum analyzer; OSC, oscilloscope; ATC, autocorrelator; ESA, electrical spectrum analyzer. \textbf{b} Optical spectral evolution versus feedback strength, showing the broadband envelope (left) and the high-resolution longitudinal modes (right) sharing a common power scale; the envelope stays wide and continuous and the modes show only a slight redshift without abrupt mode disappearance. \textbf{c} Autocorrelation evolution versus feedback strength, with periodic peaks spaced by about 40~ps preserved throughout. \textbf{d} Narrow-span RF power map around the fundamental beatnote versus feedback strength, showing a transition into a feedback-stabilized regime with external-cavity-related side features spaced by $\Delta f_\mathrm{ext}\approx11$~MHz. \textbf{e} Fundamental RF beatnote in the free-running state and under $-10$~dB feedback, both displayed over a 10~MHz span at 500~Hz resolution bandwidth. The free-running linewidth of 190~kHz is fitted from this span, whereas the feedback linewidth of 4.50~kHz is fitted from a separate 1~MHz-span, 100~Hz-resolution measurement that resolves the narrowed beatnote, giving a $\sim$42-fold reduction.}
\label{fig:feedback}
\end{figure*}

\subsection{Isolator-free transmission from the broadband regime}
\label{sec:transmission}

Having identified the high-injection flat-top branch, we next tested whether its lines are usable communication carriers rather than merely components of a broad optical spectrum. Isolator-free intensity-modulation/direct-detection (IM/DD) transmission was performed one comb line at a time, using the setup of Fig.~\ref{fig:transmission}a. The laser was biased in the broadband-comb regime, and no optical isolator was placed between the laser facet and the transmission link. Each selected carrier was externally modulated with a 25~Gb/s non-return-to-zero (NRZ) signal and detected for eye-diagram and bit-error-rate (BER) measurement. The full link is described in Materials and methods.

Figure~\ref{fig:transmission}b shows the comb-line spectrum used as wavelength carriers (upper) and the corresponding per-channel BER at 25~Gb/s NRZ (lower), each point taken at the maximum received optical power of that channel ($-5.9$ to $-6.9$~dBm). All 94 carriers within the 13.73~nm 3~dB bandwidth stay below the 7\% HD-FEC threshold of $3.8\times 10^{-3}$, with BER values typically between $10^{-6}$ and $10^{-4}$ across the central wavelengths and rising mildly toward the band edges. The absence of channel-to-channel collapse, despite operation without an optical isolator, indicates that the broadband flat-top comb preserves sufficiently uniform carrier quality across the measured bandwidth for error-correctable multi-wavelength modulation under residual back-reflections.

A subset of representative channels (tone 1 to tone 6) distributed across the 3~dB bandwidth was selected for detailed eye-diagram and power-sensitivity characterization. Figure~\ref{fig:transmission}c shows the back-to-back eye diagrams, with clear and uniformly open eyes across the selected wavelengths. The corresponding BER versus received optical power, Fig.~\ref{fig:transmission}d, obtained by attenuating the received power below the per-channel maximum, follows the expected log-linear trend for IM/DD detection, with all channels reaching the HD-FEC threshold at comparable sensitivities and confirming consistent performance across the comb.

At the 25~GHz comb spacing, the Nyquist grid allows up to 25~Gbaud per channel without spectral overlap between adjacent carriers. The present NRZ format carries one bit per symbol for 25~Gb/s per channel. Higher aggregate capacity on this fixed grid comes from higher spectral efficiency: advanced modulation formats such as PAM-4 or PAM-8 encode more bits per symbol at the same symbol rate, enabled by suitable transmitter electronics and digital signal processing. Taking the 110 carriers available within the record 3~dB optical bandwidth of this QD MLL platform as the projection basis, the 25~Gb/s NRZ format corresponds to an aggregate line rate of $2.75$~Tb/s from a single electrically pumped device.

\subsection{Transition to a feedback-stabilized regime under controlled reflection}
\label{sec:feedback}

Finally, we evaluate the response of the 25~GHz QD MLL to optical feedback, directly relevant to isolator-free photonic links where residual reflections from facets, filters, couplers, and chip interfaces can re-enter the laser cavity. Such feedback often destabilizes conventional semiconductor lasers, whereas QD lasers are expected to be more tolerant owing to their small linewidth enhancement factor and fast carrier dynamics~\cite{shi2026lsa}. Light was returned to the laser through a calibrated passive fiber loop, shown in Fig.~\ref{fig:feedback}a, which sets an accessible feedback range of $-48$ to $-9$~dB; the loop and its calibration are described in Materials and methods. The optical spectrum, time-domain autocorrelation waveform, and RF spectrum were recorded throughout this feedback sweep. The laser was biased at the intermediate-injection branch of the broadband-comb region ($I_\mathrm{gain}=120$~mA, $V_\mathrm{SA}=-3.5$~V) of Fig.~\ref{fig:device}e, where the free-running comb has a 3~dB bandwidth of 11.09~nm.

Figure~\ref{fig:feedback}b shows the optical spectral evolution versus feedback strength, with the broadband envelope (left) and the high-resolution longitudinal modes (right) plotted on a common power scale. The envelope stays wide and continuous over the measured range, while the high-resolution map resolves individual longitudinal modes and reveals only a slight redshift, with no abrupt mode disappearance or envelope collapse. The corresponding time-domain autocorrelation evolution in Fig.~\ref{fig:feedback}c retains its periodic peaks, spaced by about 40~ps throughout the sweep, confirming that the mode-locked temporal structure is preserved at the 25~GHz repetition period.

The RF response is summarized in Fig.~\ref{fig:feedback}d, e. The narrow-span RF power map of Fig.~\ref{fig:feedback}d shows no broadband noise pedestal indicative of coherence collapse, consistent with the full 0 to 30~GHz RF spectrum that remains free of any broadband pedestal across the entire feedback range (see Supplementary Fig.~S3). Instead, beyond a feedback strength of about $-28$~dB, the beatnote transitions into a feedback-stabilized regime accompanied by external-cavity-related side features spaced by $\Delta f_\mathrm{ext}\approx11$~MHz, indicating coherent interaction between the laser cavity and the delayed feedback. This is quantified in Fig.~\ref{fig:feedback}e, which compares the fundamental beatnote in the free-running and $-10$~dB feedback states, both displayed over a 10~MHz span at 500~Hz resolution bandwidth. The fitted Voigt linewidth narrows from 190~kHz, obtained from this span, to 4.50~kHz, obtained from a separate 1~MHz-span measurement at 100~Hz resolution bandwidth that resolves the narrowed beatnote, a $\sim$42-fold reduction consistent with the well-documented robustness of QD gain media to external feedback~\cite{verolet2020jlt,dong2020Journal}.

These results identify a distinct feedback-stabilized branch of the system response: controlled delayed reflection does not produce coherence collapse and can narrow the RF beatnote while preserving broadband mode locking. Because feedback delay and phase were not independently swept, the approximately $-28$~dB transition is specific to the tested external-cavity configuration and should not be interpreted as a universal stabilization boundary. Within that scope, the result establishes the QD MLL as a feedback-compatible and potentially feedback-enhanced multi-wavelength source for isolator-free optical interconnects.

\section{Discussion}

The central contribution of this study is the experimental identification of a bias-resolved operating landscape in a deliberately simple QD MLL, rather than a collection of isolated performance records. Table~\ref{tab:benchmark} compares the measured performance with state-of-the-art 1.3~\textmu m QD mode-locked comb sources. To our knowledge, the 16.16~nm 3~dB optical bandwidth is the broadest reported among QD mode-locked comb sources, while the 41.6~fs integrated timing jitter over the ITU-T-specified 4 to 80~MHz range is the lowest reported for a high-channel-count O-band QD passively mode-locked source. Importantly, these extrema occur at distinct mapped bias points, and the transmission and feedback measurements were performed at separate operating points selected from the broadband branch. The significance is therefore not that every performance maximum is achieved simultaneously, but that short-pulse, broadband, low-noise, transmission-ready, and feedback-stabilized operation are all accessible within one physical device and can be associated with identifiable regions of its operating plane.

The deliberately simple two-section device geometry has an important evidentiary role in this interpretation. We do not claim that the two-section cavity or tapered ridge constitutes a new laser architecture. Rather, avoiding chirped gain stacks, harmonic colliding-pulse operation, and co-doping reduces additional structural degrees of freedom that could otherwise obscure the relation between the measured behavior and the underlying QD gain-absorber dynamics.

Within the present device, the narrow SA maintains a low absorber saturation energy and efficient intensity-dependent loss, whereas the widened gain section increases the gain saturation energy, suppresses gain-saturation-induced pulse broadening, and supports power extraction while preserving the gain-absorber contrast required for passive mode locking. This saturation-energy balance provides a physical basis for accessing different comb states through electrical bias. Nevertheless, the broadband state cannot be attributed to the QD gain medium alone: gain recovery, spatial hole burning, group-velocity dispersion, Kerr nonlinearity, four-wave mixing, saturable absorption, and cavity geometry all contribute to the observed dynamics. The present measurements therefore support a regime-level interpretation consistent with AM-like and FM-like comb formation, rather than a complete microscopic reconstruction of the intracavity field.

\begin{table*}[!t]
\caption{\textbf{Comparison of recent state-of-the-art O-band QD semiconductor comb sources for optical interconnects}}
\label{tab:benchmark}
\vspace{2pt}
\sffamily
\footnotesize
\setlength{\tabcolsep}{2.5pt}
\renewcommand{\arraystretch}{1.25}
\begin{tabularx}{\textwidth}{@{}
>{\raggedright\arraybackslash}p{0.085\textwidth}
>{\raggedright\arraybackslash}p{0.115\textwidth}
>{\raggedright\arraybackslash}p{0.105\textwidth}
>{\centering\arraybackslash}p{0.045\textwidth}
>{\centering\arraybackslash}p{0.09\textwidth}
>{\centering\arraybackslash}p{0.042\textwidth}
>{\raggedright\arraybackslash}p{0.165\textwidth}
>{\raggedright\arraybackslash}X
>{\centering\arraybackslash}p{0.055\textwidth}
@{}}
\toprule
\textbf{Year/Ref.} &
\textbf{Material platform} &
\textbf{Method} &
\textbf{$f_\mathrm{rep}$ (GHz)} &
\textbf{3 dB BW (nm/THz)} &
\textbf{3 dB lines} &
\textbf{RF/noise metric} &
\textbf{Transmission demo}$^{d}$ &
\textbf{Optical isolator} \\
\midrule
2019/\cite{liu2019optica} &
InAs/GaAs QD on Si &
Two-section &
20 &
6.1 / $\sim$1.14 &
58 &
1.8~kHz RF linewidth; 82.7~fs jitter &
4.1~Tb/s (32 Gbaud Nyquist PAM4) &
N/R \\
2020/\cite{pan2020pr} &
InAs/GaAs QD &
Two-section &
25.5 &
4.81$^{a}$ / $\sim$0.8 &
-- &
Thermally stable RF spacing &
N/R &
N/R \\
2022/\cite{huang2022pr} &
InAs/GaAs QD &
Fourth-harmonic CPML &
100 &
11.5 / $\sim$2.01 &
20 &
N/R &
1.6~Tb/s (40 Gbaud PAM-4) &
w \\
2023/\cite{dong2023lsa} &
InAs/GaAs QD &
Second-harmonic CPML &
60 &
12.1 / 2.2 &
$\sim$37$^{b}$ &
$<$20 to 40~kHz RF linewidth &
N/R &
N/R \\
2024/\cite{buyalo2024sr} &
InAs/GaAs QD &
Two-section &
25 &
$\sim$10.3 / $\sim$1.83$^{c}$ &
74$^{c}$ &
49.5~kHz RF linewidth &
N/R &
w \\
2025/\cite{rautert2025ofc} &
InAs/GaAs QD &
Two-section &
100 &
$\sim$13.2 / 2.31 &
24 &
N/R &
2.54~Tb/s (106 Gb/s PAM-4) &
w \\
2025/\cite{pan2025lpr} &
InAs/GaAs QD &
Second-harmonic CPML &
100 &
14.31 / $\sim$2.50 &
26 &
N/R &
3.33~Tb/s (128 Gb/s PAM-4) &
w/o \\
This work &
InAs/GaAs QD &
Two-section &
25 &
16.16 / 2.75 &
110 &
0.58~kHz RF linewidth; 41.6~fs jitter &
2.75~Tb/s (25 Gb/s NRZ) &
w/o \\
\bottomrule
\end{tabularx}
\vspace{2pt}

{\sffamily\fontsize{7.4}{9}\selectfont
$^{a}$ Reported as 6~dB bandwidth in ref.~\citenum{pan2020pr}. $^{b}$ Estimated from the reported 2.2~THz 3~dB bandwidth and 60~GHz comb spacing. $^{c}$ Estimated from the reported 74 modes within a $\pm 1.5$~dB intensity range and 25~GHz comb spacing in ref.~\citenum{buyalo2024sr}. $^{d}$ Aggregate transmission capacity or projected line rate; the modulation format used in each reference is indicated. CPML: colliding-pulse mode-locked laser; N/R: not reported; w: optical isolator used or included in the setup/package; w/o: no optical isolator used or no built-in optical isolator reported. ``--'' indicates that the metric was not reported or is not directly comparable.\par}
\end{table*}

The bias-plane maps clarify why a single ``best'' operating point is not meaningful for this platform. At low injection and strong SA reverse bias, the device produces 0.80~ps pulses, consistent with an AM-like short-pulse regime. At high injection and intermediate reverse bias, the 16.16~nm flat-top spectrum coexists with a temporally extended waveform of 14.98~ps effective width, consistent with a stronger FM-like contribution. The minimum RF timing noise occurs at a third, middle-bias operating point. For wavelength-parallel interconnects and parallel optical processing, the broadband branch is the relevant optimum because carrier count, spectral flatness, and per-line usability are more important than transform-limited pulse duration. Time-domain sampling or optical clocking instead favors the short-pulse branch, whereas microwave-photonic functions can target the low-noise branch. The operating map therefore converts an apparent trade-off among bandwidth, pulse duration, and RF coherence into bias-selectable functionality within one device.

The 25~GHz fundamental-cavity design should be viewed as a diagnostic platform rather than as the final channel-spacing choice for every DWDM transmitter. Wider spacings, such as 100~GHz, relax optical filtering and reduce modulation-induced inter-channel crosstalk for high-baud-rate links. The denser 25~GHz grid instead maximizes the number of carriers available within the O-band gain window and provides a stringent test of gain bandwidth, line uniformity, RF coherence, transmission compatibility, and feedback response. It also places the fundamental beatnote within the direct detection range of standard electrical spectrum analyzers, enabling quantitative RF-linewidth, phase-noise, and feedback-response characterization without the frequency-extension or down-conversion methods required for typical 100~GHz-class combs. The platform therefore establishes a baseline from which the same material and process can be extended toward higher-repetition-rate QD combs.

The isolator-free transmission measurements connect the broadband regime to a specific system requirement. Sequential per-line testing shows that all 94 measured carriers across the 13.73~nm transmission-bias bandwidth support 25~Gb/s NRZ transmission below the 7\% HD-FEC threshold without an optical isolator between the laser and the link. These measurements establish the usability of the individual carriers, while the 2.75~Tb/s aggregate line rate remains a projection obtained by applying the measured per-carrier rate to the 110 carriers available at the maximum-bandwidth bias. A simultaneous fully loaded WDM experiment will be required to quantify aggregate penalties arising from channel filtering, modulator bandwidth, crosstalk, and shared amplification.

The controlled-feedback experiment identifies a further operating regime relevant to isolator-free integration. Across the explored feedback range, the broadband optical envelope and 25~GHz temporal structure remain preserved, with no evidence of broadband coherence collapse. Beyond approximately $-28$~dB feedback, the RF beatnote develops external-cavity-related side features separated by approximately 11~MHz and enters a substantially narrower state, with the fitted linewidth decreasing from 190~kHz to 4.50~kHz at $-10$~dB feedback. This behavior is consistent with coherent interaction between the laser cavity and the delayed feedback path and shows that residual reflection does not necessarily destabilize the broadband comb. Under an appropriate external-cavity condition, it can instead reorganize the RF dynamics into a feedback-stabilized state. This observation is relevant to a general constraint in integrated photonics: when a practical on-chip isolator is unavailable, the design problem is not only to maximize feedback tolerance, but also to identify and control the feedback states that arise from downstream reflections.

Together, these results show that the key capabilities of an O-band interconnect comb arise from selectable regions of one gain-absorber operating landscape rather than from a single universal optimum. The contribution is therefore not a new cavity architecture or an isolated record, but a regime-resolved framework connecting QD comb-formation dynamics to bandwidth, coherence, isolator-free transmission, and feedback response. This framework establishes a quantitative baseline for scaling QD comb sources to higher repetition rates and toward terabit-scale wavelength-parallel interconnects, optical computing, and microwave photonics.

\section{Materials and methods}

\subsection{QD laser fabrication}

The laser wafer was grown by molecular beam epitaxy on an \textit{n}-type GaAs substrate. The active region contains eight periods of self-assembled InAs quantum dots in an InGaAs dot-in-a-well configuration with GaAs barriers, placed inside a GaAs waveguide core and confined by Al$_{0.4}$Ga$_{0.6}$As cladding on both sides. A 300~nm \textit{n}-GaAs layer beneath the \textit{n}-cladding serves as the back-contact layer, and heavily doped \textit{p}-GaAs layers on top form the \textit{p}-contact. The complete layer structure is listed in Supplementary Table~S1.

Fabrication used two mesa etches. A patterned PECVD SiO$_2$ hard mask defined the waveguide ridge, which was transferred into the upper cladding by a shallow inductively coupled plasma (ICP) dry etch. This shallow ridge sets the lateral optical confinement while keeping sidewall-scattering loss low. A second SiO$_2$ mask defined the wider current-injection mesa, and a deeper ICP etch exposed the \textit{n}-GaAs contact layer. An Ni/Ge/Au/Ni/Au \textit{n}-contact was deposited on the exposed \textit{n}-GaAs by electron-beam evaporation and annealed. A conformal $\sim$500~nm SiO$_2$ film was then deposited to passivate the surface and the mesa sidewalls. Isolation between the gain and SA sections was defined by a separate hard mask and an isolation etch across a 10~\textmu m gap, which removed the 300~nm heavily doped \textit{p}-contact layer together with about 300~nm of the underlying \textit{p}-cladding and separated the two \textit{p}-electrodes. Contact windows were opened through the passivation, a Ti/Pt/Au \textit{p}-contact was formed on the ridge top by lift-off, and thick Ti/Au probe pads were patterned over both sections for electrical access. Both facets were left as-cleaved, and the bars were mounted on a copper heat sink. The full process flow is illustrated in Supplementary Fig.~S1.

\subsection{Isolator-free transmission experiment}

Isolator-free intensity-modulation/direct-detection transmission was measured one comb line at a time, using the setup of Fig.~\ref{fig:transmission}a. The laser was biased in the broadband-comb regime, and no optical isolator was placed between the laser facet and the link, so that residual reflections from the downstream components could re-enter the laser cavity. A tunable optical bandpass filter selected a single comb line as the optical carrier. An O-band semiconductor optical amplifier compensated the insertion losses of the filter, the modulator, and the fiber components. A polarization controller aligned the carrier polarization to a 30~GHz Mach-Zehnder modulator, which was driven through a broadband RF amplifier by a 25~Gb/s non-return-to-zero signal from a bit-error-rate tester, with a pseudo-random binary sequence (PRBS) length of $2^{15}-1$. The modulator DC bias was adjusted slightly during the wavelength sweep to maintain near-quadrature operation. A variable optical attenuator set the received optical power, and the signal was then split into two paths: one to a digital communication analyzer for back-to-back eye diagrams, and one to a 70~GHz photodetector followed by the bit-error-rate tester for BER counting. Each channel was first recorded at its maximum received optical power, between $-5.9$ and $-6.9$~dBm, after which the received power was attenuated to obtain the BER-versus-power curves.

\subsection{Optical feedback characterization}

Figure~\ref{fig:feedback}a shows the setup used to characterize optical feedback. The laser was mounted on a thermoelectric cooling stage held at 20~$^{\circ}$C. The emission was collected by a lensed fiber and routed through a three-port fiber-optic circulator, and an optical coupler split the light into a feedback path and a monitoring path. In the feedback path, a variable optical attenuator set the feedback strength and a manual polarization controller aligned the returned polarization to the transverse-electric mode of the laser before re-injection. The accessible feedback range was set by the passive loop and component losses to $-48$ to $-9$~dB. An optical isolator in the monitoring path prevented instrument reflections from reaching the laser, so that the applied feedback was set solely by the calibrated loop.

The feedback strength is defined as the ratio of the power returned to the laser facet to the power emitted from that facet, and therefore accounts for both the forward and the backward coupling loss. It was obtained from a per-component loss calibration performed with a calibrated power meter and an integrating sphere: the on-chip output was compared with the fiber-coupled power to extract the one-way chip-to-fiber coupling loss of about 2.5~dB, the insertion losses of the circulator, coupler, and polarization controller were measured separately, and the attenuator was then scanned to set the net on-chip feedback level. At each feedback level, the optical spectrum was recorded on an optical spectrum analyzer, the temporal structure on an intensity autocorrelator and a sampling oscilloscope, and the beatnote on an electrical spectrum analyzer.

\begingroup\bmsize

\bmhead{Acknowledgements}
\noindent This work was supported by King Abdullah University of Science and Technology (KAUST) under Award No. RFS-TRG2024-6196, ORFS-CRG12-2024-6487, RFS-sTRG2026-7454, ORFS-2026-5874. The authors acknowledge the support from the Nanofabrication Core Lab at KAUST.

\bmhead{Author contributions}
\noindent Y.S. fabricated the QD laser devices with assistance from X.O., W.H., and X.Y. Y.S. and W.H. established the MLL characterization platform and performed the bias-resolved MLL characterization and analysis with assistance from D.S. Y.S. and W.L. conducted the data transmission experiments. Y.S. performed the optical feedback characterization. Y.S. and Y.W. wrote the manuscript with input from all authors. Y.W. supervised the project.

\bmhead{Data availability}
\noindent All data generated or analyzed during this study are available within the paper and its Supplementary Information. Further source data may be obtained upon reasonable request.

\bmhead{Conflict of interest}
\noindent The authors declare no conflict of interest.

\bmhead{Supplementary information}
\noindent Supplementary information accompanies this manuscript.

\endgroup

\bibliography{reference}

\end{document}